\documentclass[twocolumn, english,showpacs,preprintnumbers,amsmath,amssymb]{revtex4}
\usepackage[T1]{fontenc}
\usepackage[latin9]{inputenc}
\usepackage{verbatim}
\usepackage{amsmath}
\usepackage{amssymb}
\usepackage{graphicx}
\usepackage{esint}

\makeatletter
\@ifundefined{textcolor}{}
{%
 \definecolor{BLACK}{gray}{0}
 \definecolor{WHITE}{gray}{1}
 \definecolor{RED}{rgb}{1,0,0}
 \definecolor{GREEN}{rgb}{0,1,0}
 \definecolor{BLUE}{rgb}{0,0,1}
 \definecolor{CYAN}{cmyk}{1,0,0,0}
 \definecolor{MAGENTA}{cmyk}{0,1,0,0}
 \definecolor{YELLOW}{cmyk}{0,0,1,0}
 }

\newcounter{mycount}

\makeatother

\usepackage{babel}
\begin{document}

\title{Composite Fermion states on the torus }

\author{M. Hermanns$^{1}$}

\affiliation{$^{1}$Institute for Theoretical Physics of Cologne, 50937 Cologne,
Germany}

\date{\today}
\begin{abstract}
We extend the composite fermion construction to the torus geometry. 
We verify the validity of our construction by computing the overlap of the composite fermion state to the exact
diagonalization ground state of both Coulomb interaction and Haldane-pseudopotential interaction $V_0$ ($V_1$) for bosonic (fermionic) states. 
\end{abstract}
\maketitle

\section{Introduction}

A paramount impetus for the growing interest in strongly correlated quantum matter is the 
discovery that such systems can be topologically ordered. The first, and most prominent, examples
are the various incompressible fractional quantum Hall (FQH) liquids. These 
are formed at low temperatures, when very clean two-dimensional electron gases are subjected
to a strong perpendicular magnetic field \cite{Tsui Gossard Stoermer}.
The most striking consequence of the topological order in the FQH liquids is that the  emergent low-energy quasiparticles have fractional electric charge, 
and are believed to obey fractional braiding statistics \cite{Leinaas and Myrheim}. 
Another hallmark of a topological phase of matter is that the number of degenerate ground states depend on the topology of the space on which the state is defined. 
Again, the simplest example is the FQH liquids, where numerical solutions of the microscopic Coulomb problem can be compared to  different theoretical predictions. 
In this paper we develop techniques that make it possible to make such comparisons for the Jain states, which are an important set of actually observed FQH liquids. 

The problem of many particles moving in a strong magnetic field, and  interacting via Coulomb forces, is intractable, so one has to find an effective description.
While a  natural approach is to construct effective low-energy field theories \cite{effective theory, Lopez}, a very fruitful alternative route 
--- following a seminal paper by Laughlin \cite{Laughlin}--- has been to construct model wave functions with well defined topological  properties,  and verify numerically  that they describe small systems accurately. 
The most common method of verification is to compute the overlap with the exact Coulomb ground state, but recently it was shown that studying the entanglement spectrum \cite{LiHaldane} can  give valuable complementary insights. 

Although most effort in FQH physics during the last decade has been aimed at understanding different non-Abelian states, we shall here concentrate on the family of observed states in the lowest Landau level. There are two successful theoretical approaches to describe these states---the Haldane-Halperin hierarchy \cite{Haldane Pseudopotentials,Halperin hierarchy} and the composite fermion (CF) theory \cite{first CF paper}.

The Haldane-Halperin hierarchy describes a family of incompressible FQH liquids formed via successive condensation of the low-energy excitations--- quasiholes and quasielectrons. 
It predicts that incompressible FQH liquids may be found at filling fractions $\nu=p/q$, where $p$ and $q$ are relatively prime integers and $q$ is odd. 
It also argues that the stability  of the liquids decreases roughly as $\sim 1/q$ with increasing denominator. 
The emergent quasiparticles have fractional electric charge $1/q$ and obey fractional Abelian braiding statistics. 
The simplest way to obtain explicit wave functions for ground and quasiparticle states,  at all levels of the hierarchy, is by using  conformal field theory techniques \cite{qe hierarchy, full hierarchy}.

The CF theory can describe most of the observed FQH liquids by mapping the problem
of strongly interacting fermions that fill a fraction $\nu $ of a Landau level, to  that of non-interacting (or at least weakly interacting) composite fermions filling  an integer number of effective Landau levels in a  reduced  magnetic field.
The latter describes an incompressible state because of the finite gap between the effective Landau levels. 
This approach also gives a simple picture of the ground state and  the low-energy sector. 
A big initial success of the CF theory was the very good agreement with the exact Coulomb eigenstates obtained from finite-size numerical studies, both for the ground state and the excited states. 
The two approaches are not exclusive, but rather describe the same universal features \cite{Read} and also give very similar predictions for the relative stability of the various FQH liquids \cite{Halperin hierarchy,first CF paper}.
Moreover, it was shown that for the important case of the positive Jain series, the CF wave functions can be obtained by a hierarchical construction, both in planar and spherical geometry \cite{first CFT,long qe, bonderson, CFT wf on sphere}.

Both approaches to the Abelian FQH states in the lowest Landau level---the hierarchy and the CF theory---are well understood (and studied) in the disk and sphere geometry; see eg. \cite{Jainbook} and references therein.
The torus geometry, however, has been studied far less and explicit wave function were only known for certain model states that are determined uniquely (up to center-of-mass translations) by their vanishing properties \cite{Haldane and Rezayi Torus,MRtorus,MR qhs torus}.
A first attempt to construct hierarchical wave functions on the torus was done in  Ref. \cite{CFT torus} using conformal field theory techniques. 
While it provided wave functions that are very good approximations to the exact Coulomb ground state in a certain parameter regime, the construction was not  satisfactory in that the wave functions did not transform properly under modular transformations and were uniquely defined only in the thermodynamic limit. 
It has only  recently been understood how to resolve these problems \cite{aps talk}.

In this paper, we show how to generalize the CF construction to the torus geometry. 
The construction has no free parameters and gives unique model wave functions (up to center-of-mass translations) of the ground state and the excited states at filling fraction $\nu=\frac{n}{np+1}$, $n$ and $p$ integers, corresponding to the positive Jain series.  

There are several reasons why the torus geometry is interesting, even though it cannot be realized experimentally. We already mentioned the topological ground state degeneracy, but it is also important that 
numerical calculations are better defined on closed manifolds, such as the sphere and the torus, since they do not suffer from edge effects, which can be substantial for the system sizes one can reach numerically. 
While numerics on the sphere have proven very useful, there are still problems connected to finite size, most notably the so-called shift. 
It can happen that two states that are at the same (thermodynamic) filling fraction $\nu$ appear at different magnetic fluxes $N_{\phi}$ in the finite-size system. 
The most prominent examples are the Moore-Read state \cite{MR} and the CF Fermi liquid \cite{CF fermi liquid}, which are both at filling $1/2$ but have different shift. 
Numerical comparison of these two states on the sphere is therefore only indirect.
On the torus, this issue does not arise, which allows for a directly comparison of these states \cite{MRvsFL}. 

Another advantage of the torus geometry is that one can change the shape of the torus--- described by the modular parameter $\tau$--- and thus get more information about a state without having to increase the system size. 
This was successfully used for entanglement entropy calculations, where one wants to extract a subleading constant term in the entropy.
Ref. \cite{entanglement entropy on torus} showed that changing the aspect ratio of the torus and thus obtaining additional data, yielded much more accurate bounds on the topological constant than could be obtained from sphere calculations. Also, as shown by Avron {\it et.al.} \cite{viscosity}, by studying the response of QH liquid to an adiabatic change in $\tau$, one can determine the odd part of the viscosity tensor. An explicit  calculation in the case of  the Laughlin states was made by Read \cite{Read viscosity}, and  the results in this paper could be used to perform similar calculations for the Jain states. 
Let us also note that the techniques introduced in this paper are not restriced to the positive Jain series, but can also be used to study more exotic states, such as the Bonderson-Slingerland states \cite{BS states}, the non-Abelian condensate states \cite{NAC}, as well as the closely related bipartite CF states \cite{bipartite CF}. 

\paragraph*{Outline of the paper}
In Sec. \ref{sec:General CF construction} we first present the CF construction on the disk geometry and show how to generalize the approach to the torus geometry.
We discuss single-particle states on the torus in Sec.  \ref{sub:Single-particle-states-torus} and give an expression for the product of two such states at different magnetic fluxes $N_{\phi_{1}}$ and $N_{\phi_{2}}$ in Sec. 
\ref{sub:product single particle states}. 
The derivation of this identity is given in the Appendix. 
In Sec.  \ref{sub:CF wave function on torus} we show how to evaluate CF states on the torus. 
Overlaps of some CF states with exact diagonalization results using both Coulomb and Haldane pseudo-potential interactions are calculated in Sec.  \ref{sec:Numerical-analysis}.
These overlaps should be regarded solely as a proof of principle that the construction on the torus is sound. 
In Sec.  \ref{sec:Discussion}, we speculate on possible lowest Landau level projection schemes in real space, given that the torus places additional constraints on model wave functions.

%%%%%%%%%%%%%%%%%%%%%%%%%%%%%%%%%%%%%%%%%%%%%%%%%%%%%%%%%%%%%%%%%%%%%%%%%%%%%%%%%%%%%%%%%%%%%%%%%%%%%%%%%%%%%%%%%%%%%%%%%%%%%%%%%%%%%%%%%%%%%%%%%%%%%%%%%%%%%%%%%%%%%%%%%%%%%%%%%%%%%%%%%%%%%%%%%%%%%%%%%%%%%%%%%%%%%%%%%%%%%%%%%%%%%%%%%%%%%%%%%%%%%%%%%%%%%%%%%%%%%%%%%%%%%%%%%%%%%%%%

\section{General composite fermion construction}

\label{sec:General CF construction} In this Section we explain how to generalize the CF construction to the torus geometry. 
In Sec.  \ref{sub:CFdisc} we first discuss the CF construction on the disk and sphere and point out some subtleties that become important on the torus. Sec.  \ref{sub:Single-particle-states-torus} contains a short review on the single-particle states on the torus. 
In Sec. \ref{sub:product single particle states}, we derive formulas for the projection of a product of two single-particle states. 
In Sec. \ref{sub:CF wave function on torus}, we discuss properties of the CF states on the torus using the bosonic CF state at $\nu=2/3$ as an explicit example. 

\subsection{Composite fermions on the disk geometry\label{sub:CFdisc}}

There are already many good texts on the CF construction---see, for instance, \cite{Jainbook} for an extensive and pedagogical review. 
Thus, we keep the discussion in this Section very brief and focus on properties that are relevant for the torus. 
In the following, we restrict ourselves to the positive Jain series at fillings $\nu=\frac{n}{np+1}$ , where $n\geq1$ and $p$ are integers. 
We expect the negative Jain series to work analogously, but we have not yet performed any explicit calculations. 
In the CF theory, one attaches an even (odd) number of vortices to strongly interacting fermions (bosons).
The resulting fermionic particles are called composite fermions and one assumes that these composite particles are non-interacting or at least very weakly interacting. 
Due to the attachment of vortices, they feel a reduced magnetic field $B^{\star}=B-p\rho \phi_{0}$, where $\phi_{0}$ is the magnetic flux quantum and $\rho $ is the two-dimensional density. 
For properly chosen $p$ the reduction of magnetic flux is such that the CFs fill an integer number of effective Landau levels. 

A trial wave function for the ground state of strongly interacting
particles at filling $\nu=\frac{n}{np+1}$ is then usually written as 
\begin{align}
\Psi_{\nu}(\{z_{j}\}) & =\mathcal{P}_{LLL}\left\{ \Phi_{n}(\{z_{j},\bar{z}_{j}\})\prod_{i<j}(z_{i}-z_{j})^{p}\right\} \,,\label{eq:Jain state disc}
\end{align}
where $z=x+iy$ is a complex coordinate. 
$\Phi_{n}(\{z_{j},\bar{z}_{j}\})$ is the many-body wave function (slater determinant) for the $n$ lowest Landau levels filled and $\mathcal{P}_{LLL}$  projects to the lowest Landau level. 
Equation \eqref{eq:Jain state disc} does not strictly speaking describe a proper lowest Landau level wave function on the disk, because it does not have the correct Gaussian factor. 
Usually, one does not worry about this but just adds the correct factor by hand. 
However, this subtlety becomes important on the torus as explained in the next paragraph. 

The naive guess of how to generalize Eq. \eqref{eq:Jain state disc} is to replace each part by the respective torus counterpart. 
In particular,  this would amount to replacing  the Jastrow factor with its periodized version \cite{Haldane and Rezayi Torus}
\begin{align}
\prod_{i<j}(z_{i}-z_{j})^{p} & \rightarrow\prod_{i<j}\theta_{1}(z_{i}-z_{j}|\tau)^{p}\,,\label{eq:periodized Jastrow}
\end{align}
where $\theta_{1}$ is the odd Jacobi $\theta$ function (defined by setting $a=b=1/2$ in Eq. \eqref{eq:def theta-fct}). 
We choose $\theta_{1}$ because it is the only $\theta$ function that has the correct short-distance behavior, ie. is it the only antisymmetric $\theta$-function. 
However, this choice poses two obvious problems: first, the wave function does not obey the correct boundary conditions on the torus, see Eq. \eqref{eq:boundary condition}.
\footnote{One can of course remedy the first problem by adding the appropriate center-of-mass pieces 'by hand'. However, in contrast to the Laughlin case we do not expect the center-of-mass piece to be uniquely defined for given many-body momentum as there are no model Hamiltonians with the Jain states as unique exact ground states. } 
Second, there is no efficient way to project the many-body wave function to the lowest Landau level. 
To the best of our knowledge, no analog of the Girvin-Jach  projection  \cite{LLL projection} is  known on the torus. 
We will comment more on this in Sec.  \ref{sec:Discussion}.

Instead of  Eq. \eqref{eq:Jain state disc} we will consider the following expression:
\begin{align}
\Psi_{\nu}(\{z_{j}\}) & =\mathcal{P}_{LLL}\left\{ \Phi_{n}(\{z_{j},\bar{z}_j\})\Phi_{1}(\{z_{j}\})^{p}\right\} \,.\label{eq:Jain state general geometry}
\end{align}
The replacement of the Jastrow factor to $\Phi_1$ is of course trivial for both the disk and sphere--- in the former case it only differs by a Gaussian factor. 
The point is, however, that \eqref{eq:Jain state general geometry} is a proper Landau level wave function on the disk, i.e. it has the correct Gaussian factor because the Gaussian factors of $\Phi_n$ and $\Phi_1^p$ combine to give the correct factor at the combined flux. 
In addition, using expression \eqref{eq:Jain state general geometry} solves both problems mentioned in the previous paragraph. 
It is straightforward to verify that $\Psi_{\nu}(\{z_{j}\})$ obeys the boundary conditions  on the torus \eqref{eq:boundary condition}. 
The projection onto the lowest Landau level can be implemented on the single-particle level, which is explained in Sec. \ref{sub:product single particle states}. 

\subsection{Single-particle states on the torus\label{sub:Single-particle-states-torus}}

We consider a torus spanned by two, not necessarily orthogonal, translation
vectors $\vec{L}_{1}$ and $\vec{L}_{2}$. A homogeneous external
magnetic field--- perpendicular to the surface of the torus--- is
described in terms of the vector potential $\vec{A}=-By\hat{x}$ using
Landau gauge. The number of flux quanta piercing the torus is related
to the area $\mathcal{A}=|\vec{L}_{1}\times\vec{L}_{2}|$ of the torus
by $2\pi\ell_{B}^{2}N_{\phi}=\mathcal{A}=L_{1}L_{2}\sin(\theta)$
with magnetic length $\ell_{B}=\sqrt{\hbar c/(eB)}$ and $\theta$
being the angle between $\vec{L}_{1}$ and $\vec{L}_{2}$. The case
of a rectangular torus corresponds to $\theta=\pi/2$. The shape of
the torus is conveniently parametrized by the aspect ratio 
\begin{align}
\tau=\frac{L_{2}}{L_{1}}e^{i\theta}\,.\label{eq:tau}
\end{align}

In the presence of the magnetic field, any valid wave function on
the torus must be invariant (up to an overall phase) under single-particle
magnetic translations $t(\vec{L}_{1})$ and $t(\vec{L}_{2})$, where
the magnetic translation operator is defined as 
\begin{align}
\hat{t}(\vec{L}) & =\exp\left[\vec{L}(\vec{\nabla}-i\frac{e}{\hbar c}\vec{A})-i\frac{\vec{L}\times\vec{r}}{\ell_B^2}\right]\,.\label{eq:magnTranslDef}
\end{align}
 Let us define "small" magnetic translations 
\begin{align}
\hat{t}_{1} & \equiv\hat{t}\left(\frac{\vec{L}_{1}}{N_{\phi}}\right)  =\exp\left[\frac{L_{1}}{N_{\phi}}\partial_{x}\right]\nonumber \\
\hat{t}_{2} & \equiv\hat{t}\left(\frac{\vec{L}_{2}}{N_{\phi}}\right)  =\exp\left[i\pi\frac{L_{2}\cos\theta}{L_{1}N_{\phi}}+2\pi i\frac{x}{L_{1}}\right]\nonumber \\
&\times \exp\left[\frac{L_{2}\cos\theta}{N_{\phi}}\partial_{x}+\frac{L_{2}\sin\theta}{N_{\phi}}\partial_{y}\right]\,.\label{eq:t_1 and t_2}
\end{align}
 The periodic boundary conditions of a wave function $\psi$ can,
thus, be formulated as
\begin{align}
\hat{t}_{1}^{N_{\phi}}\psi & =e^{i\alpha_{1}}\psi\nonumber \\
\hat{t}_{2}^{N_{\phi}}\psi & =e^{i\alpha_{2}}\psi\,. \label{eq:boundary condition}
\end{align}
 In the remainder of the paper, we will set the solenoid fluxes
$\alpha_{1},\alpha_{2}=0$ without loss of generality. 

As $\hat{t}_{1}$ and $\hat{t}_{2}$ do not commute with each other,
we can choose the single-particle states to be eigenstates of only
one of them. In the following, we will mostly use eigenfunctions of
$\hat{t}_{1}$ 
\begin{multline}
\phi_{n,j}^{\ell_{B}}(x,y)  =\mathcal{N}_{n}^{\ell_{B}}\sum_{k=-\infty}^{\infty}e^{-2\pi i(j+kN_{\phi})z} e^{-y^{2}/(2\ell_{B}^{2})} \\
  \times\exp\left[\frac{i\pi\tau}{N_{\phi}}(j+kN_{\phi})^{2}\right] H_{n}\left(\frac{2\pi\ell_{B}}{L_{1}}(j+kN_{\phi})-\frac{y}{\ell_{B}}\right), \label{eq:phi}
\end{multline}
 where $z=(x+iy)/L_{1}$ is the dimensionless complex coordinate of
the particles, $n=0,1,\ldots$ is the Landau level index and $j=0,\ldots,(N_{\phi}-1)$
the momentum index. $H_n$ denotes the $n$th Hermite polynomial. 
Note that the momentum is only defined modulo
$N_{\phi}$, because any larger value can be absorbed into the sum
over windings around the torus. The normalization constant is given
by 
\begin{align}
\mathcal{N}_{n}^{\ell_{B}} & =\left(\frac{\sqrt{2N_{\phi}\Im(\tau)}}{(2^{n}n!\mathcal{A})}\right)^{1/2}\,,\label{eq:norm}
\end{align}
 where $\mathcal{A}=2\pi\ell_{B}^{2}N_{\phi}$ is the total area of
the torus and $\Im(\tau)=(L_2/L_1)\sin(\theta)$ is the imaginary part of $\tau$. 
$\phi_{n,j}^{\ell_B}$ is an eigenfunction of $\hat{t}_{1}$ with
eigenvalue $\exp[-2\pi ij/N_{\phi}]$, while $\hat{t}_{2}$ shifts
the momentum by 1: $\hat{t}_{2}\phi^{\ell_B}_{n,j}=\phi^{\ell_B}_{n,j-1}$. As we
will need to distinguish single-particle states at different flux
later on, we keep the magnetic length $\ell_{B}$ as an explicit parameter
in the single-particle state $\phi_{n,j}^{\ell_{B}}$.

\subsection{Product of single-particle states on the torus\label{sub:product single particle states}}

In complete analogy to the disk and sphere geometry, we can write
a product of two single-particle states on the torus at magnetic flux
$N_{\phi_{1}}$ and $N_{\phi_{2}}$ as 
\begin{multline}
\phi_{n_{1},j_{1}}^{\ell_{1}}(x,y)\phi_{n_{2},j_{2}}^{\ell_{2}}(x,y) \\ =\sum_{n=0}^{n_{1}+n_{2}}\sum_{j=0}^{N_{\phi}-1}C_{j_{1},j_{2};j}^{n_{1},n_{2};n}\phi_{n,j}^{\ell}(x,y)\, ,\label{eq:general Clebsch}
\end{multline}
where the magnetic lengths are related by $\ell^{-2}=\ell_{1}^{-2}+\ell_{2}^{-2}$,  which is equivalent to $N_{\phi}=N_{\phi_{1}}+N_{\phi_{2}}$. The
constants $C_{j_{1},j_{2};j}^{n_{1},n_{2};n}=C_{j_{1},j_{2};j}^{n_{1},n_{2};n}(N_{\phi_{1}},N_{\phi_{2}},\tau)$
depend on the fluxes $N_{\phi_{1,2}}$ as well as the aspect ratio
of the torus. They can be computed for arbitrary $n_{1}$ and $n_{2}$,
but we have not  been able to find a closed formula except for the
two simplest cases $(n_1,n_2)=(0,0)$ and $(1,0)$. 
For the CF construction we need to know the coefficients
for $n_{2}=n=0$, but arbitrary $n_{1}$. In the following, we restrict
ourselves to these cases. 

In order to simplify notation later on we define $Q$ as the greatest
common divisor $(gcd)$ of $N_{\phi_{1}}$ and $N_{\phi_{2}}$:
\begin{align}
Q= & gcd(N_{\phi_{1}},N_{\phi_{2}})\nonumber \\
N_{\phi_{1}} & =t_{1} Q\nonumber \\
N_{\phi_{2}} & =t_{2} Q\nonumber \\
N_{\phi} & =(t_{1}+t_{2}) Q\equiv t Q\,.\label{eq:def of Q}
\end{align}
 It follows that $Q=gcd(N_{\phi_{1}},N_{\phi})=gcd(N_{\phi_{2}},N_{\phi})$.
The different magnetic lengths are related to $t_{1}$ and $t_{2}$
by:
\begin{align}
\frac{\ell}{\ell_{1}} & =\sqrt{\frac{t_{1}}{t}}\nonumber \\
\frac{\ell}{\ell_{2}} & =\sqrt{\frac{t_{2}}{t}}\,.\label{eq:relation l and t}
\end{align}

For $n_{1}=0,1$ the coefficients in \eqref{eq:general Clebsch} become
rather simple:
\begin{widetext}
\begin{align}
C_{j_{1},j_{2};j}^{0,0;0} & =\sqrt{\frac{\sqrt{2\Im(\tau)}}{\mathcal{A}(-i\tau)\sqrt{Qt^{3}t_{1}t_{2}}}}\theta_{3}\left(\frac{\pi(t_{2}j_{1}-t_{1}j_{2}+\beta t_{1}t_{2}Q)}{t_{1}t_{2}N_\phi}\Big|\exp\left[\frac{\pi}{i\tau t_{1}t_{2}N_\phi }\right]\right)\label{eq:Clebsch000}\\
C_{j_{1},j_{2};j}^{1,0;0} & =-\sqrt{\frac{\pi\sqrt{2\Im(\tau)^{3}}}{(-i\tau)^{3}\mathcal{A}\sqrt{Q^{3}t^{7}t_{1}^{3}t_{2}}}}\theta'_{3}\left(\frac{\pi(t_{2}j_{1}-t_{1}j_{2}+\beta t_{1}t_{2}Q)}{t_{1}t_{2}N_\phi }\Big|\exp\left[\frac{\pi}{i\tau t_{1}t_{2}N_\phi }\right]\right)\, ,\label{eq:Clebsch100}
\end{align}
\end{widetext}
 where $j=(j_{1}+j_{2}+\beta Qt_{1})$ mod $N_\phi$, for $\beta=0,1,\ldots,t-1$. 
$C_{j_{1},j_{2};j}^{0,0;0} =C_{j_{1},j_{2};j}^{1,0;0}=0$ for $j\neq (j_{1}+j_{2}+\beta Qt_{1})$ mod $N_\phi$.  
The
third $\theta$ function is defined by:
\begin{align}
\theta_{3}(z|q) & =\theta\left[{0\atop 0}\right](z|q)\nonumber \\
\theta\left[{a\atop b}\right](z|q) & =\sum_{k=-\infty}^{\infty}q^{(k+a)^{2}}e^{2i(k+a)(z+b)}\label{eq:def theta-fct}
\end{align}
 and $\theta_{3}'(z|q)=\partial_{z}\theta_{3}(z|q)$. 
 For higher $n_{1}$, the coefficients can in principle still be represented with help of higher derivatives of the $\theta_{3}$-function, but they become increasingly cumbersome to evaluate. 
 They can be written as:
\begin{widetext}
\begin{align}
C_{j_{1},j_{2};j}^{n_{1}0;0} & =\frac{\mathcal{N}_{n_{1}}^{\ell_{1}}\mathcal{N}_{0}^{\ell_{2}}}{\mathcal{N}_{0}^{\ell}}\sum_{\beta=0}^{t-1}\delta_{j,(j_{1}+j_{2}+\beta Qt_{1})\mbox{mod }N_{\phi}}\sum_{i=0}^{\lfloor n_{1}/2\rfloor}\binom{n_{1}}{2i}\frac{(2i)!}{i!}\left(-\frac{t_{2}}{t}\right)^{i} \left(-4\pi t_{1}t_{2}\sqrt{\frac{\Im(\tau)Q}{2\pi t_1}}\right)^{n_{1}-2i}\ \nonumber \\
 & \times \sum_{s=-\infty}^{\infty}\left(s-\frac{t_{2}j_{1}-t_{1}j_{2}+\beta t_{1}t_{2}Q}{t_{1}t_{2}N_{\phi}}\right)^{n_1-2i} \exp\left[i\pi\tau t_{1}t_{2}N_{\phi}\left(s-\frac{t_{2}j_{1}-t_{1}j_{2}+\beta t_{1}t_{2}Q}{t_{1}t_{2}N_{\phi}}\right)^{2}\right]\,. \label{eq: Clebsch Gordan m00}
\end{align}
\end{widetext}
The derivation of Eq. \eqref{eq: Clebsch Gordan m00} involves straightforward but tedious algebra, which is done in the Appendix.
Equations  \eqref{eq:Clebsch000} and  \eqref{eq:Clebsch100} can be obtained from  \eqref{eq: Clebsch Gordan m00} by a Poisson resummation. 
 Note that for $\Im(\tau)$ not too small, the summation over $s$ converges rapidly. 
 For numerical purposes, one needs to care only about the first few terms around zero. 
 The problem of convergence for small $\Im(\tau)$ can in principle be avoided by doing a Poisson resummation on the sum over $s$---similar to what was done in obtaining Eqs. \eqref{eq:Clebsch000} and \eqref{eq:Clebsch100}.

\subsection{Composite fermion wave functions on the torus\label{sub:CF wave function on torus}}
The formulas derived in the previous Section allow for evaluation of the Jain state at filling $\nu=\frac{n}{np+1}$ given by
\begin{align}
\Psi_{\nu}(\{z_{j}\}) & =\mathcal{P}_{LLL}\left[\Phi_{n}(\{z_{j}\}) \Phi_{1}(\{z_{j}\})^{p}\right]\,,\label{eq:generic CF ground state}
\end{align}
 where $\Phi_{j}$ is the many-body wave function of the lowest $j$
Landau levels completely filled. For $p$ even(odd), this describes
a fermionic (bosonic) state.
As a sanity check, one may note that $n=1$ reproduces the Laughlin
states at filling $\frac{1}{p+1}$. Expression \eqref{eq:generic CF ground state}
can also be used to evaluate wave functions corresponding to quasihole
and/or quasielectron excitations--- in the same way as on the sphere
or the disk. For the sake of simplicity, we focus on ground-state wave
functions in the following discussion. 

In principle, it is straightforward to evaluate \eqref{eq:generic CF ground state},
by multiplying out the slater determinants and using the coefficients
\eqref{eq: Clebsch Gordan m00} (repeatedly if $p>1$) to reduce the
expression to a lowest Landau level wave function at the combined
flux. For the simplest state, describing the bosonic Jain state at
filling $\nu=2/3$, the explicit expression becomes:
\begin{widetext}
\begin{align}
\Psi_{2/3}(\{x_{i},y_{i}\}) & =
\sum_{\sigma\in S_{N}}(-1)^{\sigma}\left(\prod_{\alpha=0}^{N-1}\sum_{\beta_{\alpha}=0}^{2}\right)\prod_{\alpha=0}^{N/2-1}\left(C_{\alpha,\sigma(\alpha);j_{\alpha}}^{10;0}C_{\alpha,\sigma(\alpha+N/2);j_{\alpha+N/2}}^{00;0}\right)
\sqrt{N!\prod_{j=0}^{N_\phi-1}n_j!}\, m_\mu(\{z_j\})\,,
\label{eq:psi2/3}
\end{align}
\end{widetext}
where $j_{\alpha}=(\alpha\mbox{\mbox{ mod}}N_{\phi_1}+\sigma(\alpha)+\beta_{\alpha}Qt_{1})\mbox{ mod }N_{\phi}$.
The $m_\mu(\{z_j\})$'s are the many-body basis states on the torus: 
\begin{align}
 m_\mu(\{z_j\})& =\langle z_{1},\ldots,z_{N}|\mu \rangle\,,\label{eq:monomial basis on torus}
 %% =\sqrt{N!\prod_{j=0 }^{N_\phi} n_j!}^{-1}\sum_{\phi^{\ell}_\sigma\in S_N} 
\end{align}
with $\mu=\{n_0,\ldots,n_{N_\phi-1}\}$ and $n_{j}=\sum_{\alpha=0}^{N_\phi-1}\delta_{j,j_{\alpha}}$ being the
occupation number of the single-particle orbital with $\hat{t}_{1}$-eigenvalue
$\exp[-2\pi ij/N_{\phi}]$. The state $m_\mu(\{z_j\})$ is as usual defined by the properly normalized slater determinant (for fermions) or permanent (for bosons) of all the occupied single-particle states $\phi_j^\ell (x,y)$, see Eq. \eqref{eq:phi}. 

The qualitative difference between Eq. \eqref{eq:psi2/3} to the corresponding disk and sphere expressions lies in the additional sums over $\beta_1,\ldots, \beta_N$. 
In the disk and sphere geometry, the momentum  of the product of two single-particle states on the torus is simply the sum of the two momenta. 
The projection, thus, involves evaluating $N!$ terms in order to obtain the coefficients in the occupation number basis.  
On the torus, the momentum is only defined modulo the flux $N_\phi$. 
This implies  that the winding sums in the single-particle states at flux $N_{\phi_1}$ and $N_{\phi_2}$ yield different momenta at the final flux. 
For instance, in order to compute the CF state at filling fraction $\nu=2/3$, one needs to evaluate $N!3^N$ terms, which limits the system sizes one can reach. 
Note that this limitation becomes worse, if we increase $p$ in Eq. \eqref{eq:generic CF ground state}. 

In addition, \eqref{eq:generic CF ground state} is in general not an eigenstate of the many-body translation operator $\hat{T}_{1}=\prod_{j=0}^{N-1}\hat{t}_{1}^{(j)}$, where $\hat{t}_1^{(j)}$ translates the $j$th particle by $\vec L_1/N_\phi$. 
An eigenstate can be obtained by either restricting to the correct momentum sector in the Fock basis or by applying the appropriate projection operator. 
As the system is translational invariant, we expect that we can write each momentum eigenstate as a product of a wave function $\psi^{rel}$ that depends only on the relative coordinates and a wave function $\psi_j^{com}$ that only depends on the  center-of-mass coordinate $Z=\sum_{j=1}^N z_j$ and incorporates the action of the many-body translation operators $\hat T_{a}=\prod_{j=0}^{N-1}\hat t_{a}^{(j)}$, $a=1,2$. Thus, Eq. \eqref{eq:psi2/3} can be written as 
\begin{align}
\Psi_{2/3}(\{x_{i},y_{i}\}) & =\sum_{j=0}^2 c_j \psi^{rel}(\{z_j\}) \psi_{j}^{com}(Z), 
\label{eq:com}
\end{align}
where $j$ labels the many-body momentum and $c_j$ are coefficients that may depend on $\tau$ and $N$. 
However, obtaining the explicit form of Eq. \eqref{eq:com} from the Fock decomposition is a very hard, unsolved problem, because of the infinite sums appearing in the $\theta$ functions. 
For the Laughlin states--- or more generally the Read-Rezayi series--- one way to get around this problem is by guessing the correct form of $\phi^{rel}$ and using the boundary conditions \eqref{eq:boundary condition} to derive $\psi^{com}$. This is possible,  because the Laughlin state is the unique ground state of a model Hamiltonian. Unfortunately, this is not true for general CF states, which is why the decomposition into $\psi^{rel}$ and $\psi^{com}$ is not known in these cases. 

%%%%%%%%%%%%%%%%%%%%%%%%%%%%%%%%%%%%%%%%%%%%%%%%%%%%%%%%%%%%%%%%%%%%%%%%%%%%%%%%%%%%%%%%%%%%%%%%%%%%%%%%%%%%%%%%%%%%%%%%%%%%%%%%%%%%%%%%%%%%%%%%%%%%%%%%%%%%%%%%%%%%%%%%%%%%%%%%%%%%%%%%%%%%%%%%%%%%%%%%%%%%%%%%%%%%%%%%%%%%%%%%%%%%%%%%%%%%%%%%%%%%%%%%%%%%%%%%%%%%%%%%%%%%%%%%%%%%%%%%

\section{Numerical analysis\label{sec:Numerical-analysis}}

\begin{center}
\begin{figure}
\includegraphics[scale=0.37]{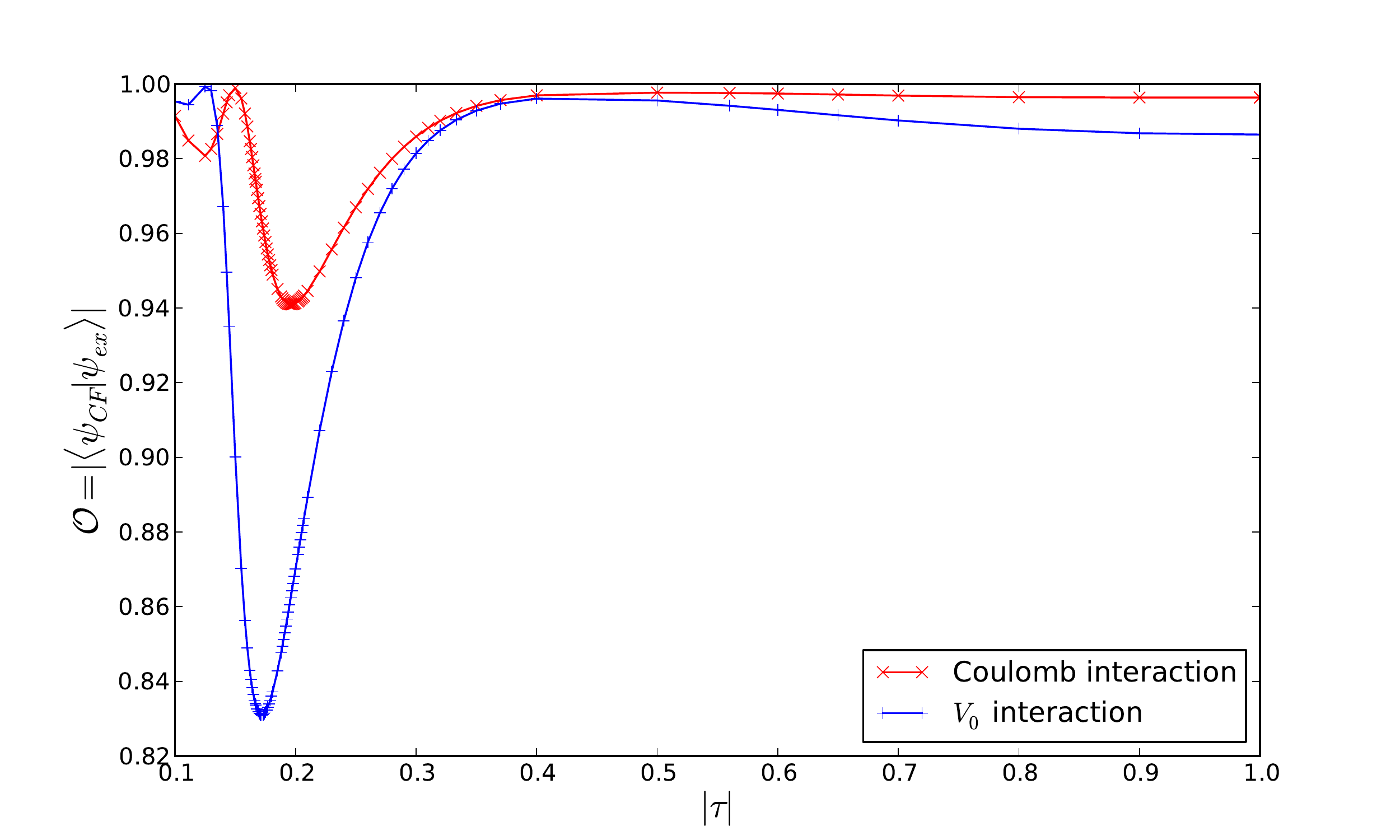}

\caption{color online: Overlap of CF state at filling $\nu=2/3$ for $N=10$
bosons with the exact diagonalization state using Coulomb interaction
(red $\times$) and a contact interaction using the Haldane pseudopotential
$V_{0}$ (blue $+$). The lines are a guide to the eye. \label{fig:bosonic overlap}}
\end{figure}

\begin{figure}
\includegraphics[scale=0.37]{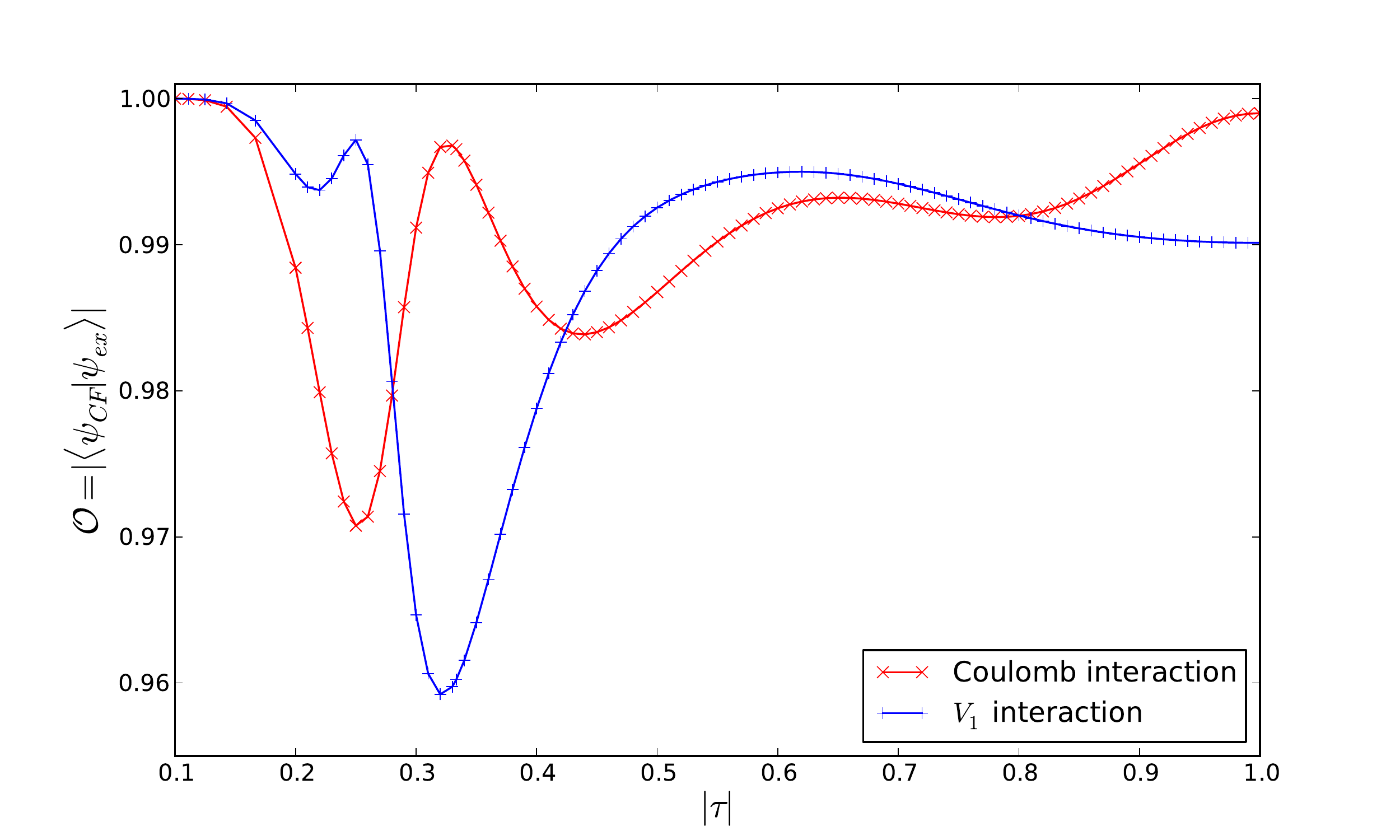}

\caption{color online: Overlap of CF state at filling $\nu=2/5$ for $N=6$
electrons with the exact diagonalization ground state using Coulomb
interaction (red $\times$) and a short-range interaction using the
Haldane pseudopotential $V_{1}$ (blue $+$). The lines are a guide
to the eye. \label{fig: fermionic overlap}}
\end{figure}

\par\end{center}
In this Section, we present numerical checks on the wave functions obtained by Eq. \eqref{eq:generic CF ground state}. 
We computed the overlap between the exact diagonalization ground state and the  bosonic CF state at filling $\nu=2/3$ for system sizes up to $n=10$ particles and in the fermionic case at filling $\nu=2/5$ for system sizes up to $N=6$ particles. 
The exact diagonalization was done both for Coulomb interaction and the smallest relevant Haldane pseudopotential \cite{Haldane Pseudopotentials}---$V_0$ for bosons and $V_1$ for fermions. 
The shape of the torus was kept rectangular ($\theta=\frac{\pi}{2}$ in Eq. \eqref{eq:tau}), with aspect ratios $|\tau|$ varying from 0.1 to 10. 
Due to the invariance of the shape of the torus under the modular transformation $\mathcal{S}:\tau\rightarrow-1/\tau$, we can restrict the analysis to aspect ratios $0.1\leq|\tau|\leq1$ without loss of generality, as a torus with $|\tau|>1$ can be obtained by an $\mathcal{S}$-transformation. 
The absolute value of the overlap $\mathcal{O}=|\langle\Psi_{CF}|\Psi_{ex}\rangle|$ between the CF state and the exact diagonalization ground state is shown in
Figs. \ref{fig:bosonic overlap} and \ref{fig: fermionic overlap} for varying aspect ratios. 

When comparing overlaps of the torus and sphere geometry, we choose the most isotropic point, namely the square torus with $|\tau|=1$. 
In the bosonic case, we find an overlap of $\mathcal{O}=0.996$ (Coulomb interaction) and $\mathcal{O}=0.973$ ($V_{0}$ interaction) for a square torus, which are slightly higher than the overlaps found in Ref. \cite{composite bosons on sphere} for the spherical geometry. 
In the fermionic case, we find overlaps $\mathcal{O}=0.999$ (Coulomb interaction) and $\mathcal{O}=0.990$ ($V_{1}$ interaction), which are slightly lower, but still comparable to the overlaps found in Ref. \cite{CF 2/5 sphere}.

The overlaps depend strongly on the shape of the torus, even though they remain quite high throughout the whole range of aspect ratios.
On general grounds, we expect the overlap to approach unity in the limit of $|\tau|\rightarrow 0$ and $|\tau|\rightarrow\infty$. 
In Ref.  \cite{thin torus long paper} it was shown that the ground state of Hamiltonians with quite generic repulsive interactions becomes a product state for aspect ratios $|\tau|\rightarrow \infty$---the so-called thin torus limit. 
We checked numerically that the CF state has this property as well.  
The other limit $|\tau|\rightarrow 0$ can---for a rectangular torus--- be mapped to the thin torus limit, when using the Landau gauge $\vec{A}=Bx\hat{y}$ and eigenfunctions of the $\hat{t}_{2}$ operator.

For all system sizes, we observe a dip (in the case of fermions several dips) in the overlap curve. 
The position of these dips depends on the system size--- in the bosonic case, it seems to be shifted to lower values of $|\tau|$ for increasing system size. 
In the fermionic case, there is too little numerical data to make a statement. 
The origin of the dips is not clear at the moment, but one can note that the overlap of the fermionic Laughlin state at $\nu=1/3$ with the Coulomb ground state has a qualitatively similar behavior as a function of $|\tau|$ to the one shown in Figs. \ref{fig: fermionic overlap}.

%%%%%%%%%%%%%%%%%%%%%%%%%%%%%%%%%%%%%%%%%%%%%%%%%%%%%%%%%%%%%%%%%%%%%%%%%%%%%%%%%%%%%%%%%%%%%%%%%%%%%%%%%%%%%%%%%%%%%%%%%%%%%%%%%%%%%%%%%%%%%%%%%%%%%%%%%%%%%%%%%%%%%%%%%%%%%%%%%%%%%%%%%%%%%%%%%%%%%%%%%%%%%%%%%%%%%%%%%%%%%%%%%%%%%%%%%%%%%%%%%%%%%%%%%%%%%%%%%%%%%%%%%%%%%%%%%%%%%%%%

\section{ Projection schemes in real space\label{sec:Discussion}}

In principle, the method described in this paper allows one to compute any CF states on the torus.  
However, evaluating Eq. \eqref{eq:generic CF ground state} becomes numerically hard for large $p$ when using Eq. \eqref{eq:general Clebsch}. 
On the torus, one needs to evaluate $(N!)^p t^N$ terms  to obtain the wave function in Fock space. 
Even on the disk and the sphere, where one does not have the additional complication of the $\beta_1,\ldots, \beta_N$ sums, the number of terms, which one needs to evaluate, is still $(N!)^p$.
This restricts the system size to very small systems for large $p$. 

A way around this, at least on the disk and the sphere, is to evaluate Eq. \eqref{eq:generic CF ground state} in real space and use Monte Carlo techniques to study the resulting model wave functions. 
How to write the projection operator $\mathcal{P}_{LLL}$  in real space was shown by Girvin and Jach in Ref. \cite{LLL projection}. 
It amounts  to moving all anti-holomorphic components to the left and replacing them by derivative operators $\bar{z}\rightarrow2\frac{\partial}{\partial z}$ with the assumption that the derivatives do not act on the Gaussian factors.

This simple implementation of the projection has no straightforward analog on the torus. 
Note that derivatives are not valid operators on the torus, because they destroy the periodic boundary conditions \eqref{eq:boundary condition}.
Following Ref. \cite{CFT torus} one may argue that the torus analog to the derivative operator should be related to the small translation operators $\hat{t}_{1}$ and $\hat{t}_{2}$ \eqref{eq:t_1 and t_2}, because the small translation operators keep the boundary conditions intact and become effectively holomorphic derivatives in the limit $L_1,L_2,N_\phi\rightarrow \infty$. 
In fact, the analog of the derivative operators may be given by a sum of translation operators that have good modular properties \cite{aps talk}. 
It is  an open question whether or not this idea could be used also to simplify the evaluation of the CF torus wave functions. 
As a modular invariant sum necessarily involves $N_{\phi}^{2}$ terms \cite{Haldane torus}, this projection scheme would be numerically cheaper than the one used here only if the sum over different translations converge rapidly.

A numerically very efficient way to evaluate  Eq. \eqref{eq:generic CF ground state} approximately is the Jain-Kamilla projection \cite{Jain Kamilla projection 1,Jain Kamilla projection 2}.
The Jain-Kamilla projection is a close approximation to the exact projection, but can--- in contrast to the latter--- be evaluated for very large system sizes. 
It  amounts to dividing the Jastrow factor in Eq. \eqref{eq:Jain state disc} as
\begin{align}
\prod_{i<j}(z_i -z_j)^{2p}=(-1)^{N(N-1)/2} \prod_{i\neq j}(z_i-z_j)^p 
\end{align}
and multiplying each column $\alpha$ of the slater determinant $ \Phi_{n}(\{z_{j}\})$ by $\prod_{j\neq \alpha}(z_\alpha-z_j)^p$. 
Then, each component of the slater determinant is projected separately to the lowest Landau level  by 
\begin{align}
\partial_\alpha  \prod_{j\neq \alpha}(z_\alpha-z_j)^p &=\sum_{j\neq \alpha}\frac{ p}{z_\alpha-z_j}\prod_{j\neq\alpha}(z_\alpha-z_j)^p\,.\label{eq:Jain-Kamilla}
\end{align}
When trying to generalize this scheme to the torus, one may note that the effect of the derivative operators in \eqref{eq:Jain-Kamilla} is to remove zeros between particles. 
On the torus, we cannot remove zeros but we may shift them.
This is in fact exactly, what the translation operators do, that were mentioned in the previous section.
However, one needs to shift the zeros without destroying the boundary conditions. 
The most straightforward implementation of \eqref{eq:Jain-Kamilla} would, thus,  require a doubly periodic function with only a single pole, which--- as we know from complex analysis--- does not exist.
Unfortunately, the boundary conditions impose rather strict rules on how one may change the wave function, which is why we have not been able to find an analog of the Jain-Kamilla projection on the torus. 

%%%%%%%%%%%%%%%%%%%%%%%%%%%%%%%%%%%%%%%%%%%%%%%%%%%%%%%%%%%%%%%%%%%%%%%%%%%%%%%%%%%%%%%%%%%%%%%%%%%%%%%%%%%%%%%%%%%%%%%%%%%%%%%%%%%%%%%%%%%%%%%%%%%%%%%%%%%%%%%%%%%%%%%%%%%%%%%%%%%%%%%%%%%%%%%%%%%%%%%%%%%%%%%%%%%%%%%%%%%%%%%%%%%%%%%%%%%%%%%%%%%%%%%%%%%%%%%%%%%%%%%%%%%%%%%%%%%%%%%%

\section{ Summary \label{sec:summary}}
In this paperarti, we generalized the CF theory to the torus geometry. 
We showed the validity of our method by calculating the overlap between the CF states and the exact diagonalization ground state of Coulomb and the smallest relevant Haldane-pseudopotential interactions for filling fractions $\nu=2/3$ and $\nu=2/5$ and system  sizes up to $N=10\,(6)$ particles for the bosonic (fermionic) states. 
The overlaps on the square torus are comparable to the ones obtained in the disk and sphere geometry. 
It turns out that numerical evaluation of the wave function is harder  on the torus than on the disk and sphere, because the winding sums mix different momentum sectors. 
We have also speculated on possible generalizations of the real space projection schemes to the torus geometry, though, unfortunately, we have not been able to find an explicit realization.
Such schemes may allow one to reach larger system sizes than are possible with the method presented here.

Let us emphasize again that our method works for the whole low-energy sector of the CF states, even though we only treated the ground states explicitly in this paper. 
The techniques introduced here may be useful for systems that cannot be studied directly on the sphere, because they have different shifts, as eg. the one studied in Ref. \cite{332 vs Jain}. 
They can also be used to study generalizations of the Abelian Haldane-Halperin hierarchy, such as the Bonderson-Slingerland states \cite{BS states}, the non-Abelian condensate state \cite{NAC}, or the bipartite CF states \cite{bipartite CF}. 
Also it will clearly be interesting to see whether the exact agreement between the hierarchy and the CF wave functions that have been demonstrated on the plane and on the sphere, also holds true on the torus.

\paragraph*{Acknowledgements}

The author wants to thank  S. Simon, M. Stone, and N. Regnault for interesting discussions and H. Hansson for interesting discussions and a critical reading of the paper. 
The exact diagonalization ground states were computed using DiagHam. 
The author wants to thank all contributors to the code, especially N. Regnault. 

%%%%%%%%%%%%%%%%%%%%%%%%%%%%%%%%%%%%%%%%%%%%%%%%%%%%%%%%%%%%%%%%%%%%%%%%%%%%%%%%%%%%%%%%%%%%%%%%%%%%%%%%%%%%%%%%%%%%%%%%%%%%%%%%%%%%%%%%%%%%%%%%%%%%%%%%%%%%%%%%%%%%%%%%%%%%%%%%%%%%%%%%%%%%%%%%%%%%%%%%%%%%%%%%%%%%%%%%%%%%%%%%%%%%%%%%%%%%%%%%%%%%%%%%%%%%%%%%%%%%%%%%%%%%%%%%%%%%%%%%%%%%%%%%%%%%%%%%%%%%%%%%%%%%%%%%%%%%%%%%%%%%%%%%%%%%%%%%%%%%%%%%%%%%%%%%%%%%%%%%%%%%%%%%%%%%%%%%%%%%%%%%%%%%%%%%%%%%%%%%%%%%%%%%%%%%%%%%%%%%%%%%%%%%%%%%%%%%%%%%%%%%%%%%

\appendix

\section{Derivation of product formula\label{app:Derivation of product formula} }

In this Appendix, we derive formula \eqref{eq: Clebsch Gordan m00}
for an expansion of the product of two single-particle states at fluxes
$N_{\phi_{1}}$ and $N_{\phi_{2}}$ in terms of the single-particle
states at the combined flux $N_{\phi}$ with $N_{\phi_{1}}+N_{\phi_{2}}=N_{\phi}$.
The product of two single-particle states is given by:
\begin{widetext}
\begin{align}
\phi_{n_{1},j_{1}}^{\ell_1}(x,y) \phi_{0,j_{2}}^{\ell_2}(x,y) & =\mathcal{N}_{n_{1}}\mathcal{N}_{0}e^{-y^{2}/(2\ell^{2})}\sum_{k_{1}=-\infty}^{\infty}\sum_{k_{2}=-\infty}^{\infty}e^{-2\pi i(j_{1}+k_{1}N_{\phi_{1}}+j_{2}+k_{2}N_{\phi_{2}})z}\nonumber \\
 & \times\exp\left[i\pi\tau\left(\frac{(j_{1}+k_{1}N_{\phi_{1}})^{2}}{N_{\phi_{1}}}+\frac{(j_{2}+k_{2}N_{\phi_{2}})^{2}}{N_{\phi_{2}}}\right)\right] H_{n_{1}}\left(\frac{2\pi\ell_{1}}{L_{1}}(j_{1}+k_{1}N_{\phi_{1}})-\frac{y}{\ell_{1}}\right)\,.\label{eq:app prod of phi's}
\end{align}
\end{widetext}
 In the following, subscript 1 and 2 denote quantities of the two
single-particle states respectively, while those without subscript
denote those of the product. Note that $N_{\phi}=N_{\phi_{1}}+N_{\phi_{2}}$
and $\ell^{-2}=\ell_{1}^{-2}+\ell_{2}^{-2}$. Let us assume $n_{2}=0$
but $n_{1}$ may be arbitrary for the time being. 
We could also consider arbitrary $n_{2}$, but it will
only complicate things unnecessarily.

Define $Q$ as the greatest common divisor $(gcd)$ of $N_{\phi_{1}}$
and $N_{\phi_{2}}$: 
\begin{align}
Q= & gcd(N_{\phi_{1}},N_{\phi_{2}})\nonumber \\
N_{\phi_{1}} & =t_{1}  Q\nonumber \\
N_{\phi_{2}} & =t_{2}  Q\nonumber \\
N_{\phi} & =(t_{1}+t_{2})  Q\equiv t  Q\,.
\end{align}
 It follows that $Q=gcd(N_{\phi_{1}},N_{\phi})=gcd(N_{\phi_{2}},N_{\phi})$.
We use that the different magnetic lengths are related as : 
\begin{align}
\frac{\ell}{\ell_{1}} & =\sqrt{\frac{t_{1}}{t}}\nonumber \\
\frac{\ell}{\ell_{2}} & =\sqrt{\frac{t_{2}}{t}}\,.\label{eq:gamma}
\end{align}
 We define the torus as in Sec. \ref{sec:General CF construction}. 
 It is easy to check that the product of the two single-particle states obeys the correct boundary conditions for flux $N_{\phi}$. 

Let us first discuss how to rewrite the double sum over windings
coming from both single-particle states, denoted by $k_{1}$ and $k_{2}$.
We can choose integers $k,\: s\in\mathbb{N}$, and $\beta\in\{0,\ldots,t-1\}$
such that 
\begin{align}
k_{1} & =\beta+k-t_{2}s\nonumber \\
k_{2} & =k+t_{1}s,
\end{align}
 which implies 
\begin{align}
t  k & =t_{1}k_{1}+t_{2}k_{2}+\beta\nonumber \\
t  s & =\beta+k_{2}-k_{1}\,.
\end{align}
It is beneficial to introduce some more notation that will simplify
expressions later on. We find that we can rewrite the phase factors
of the single-particle states as 
\begin{align}
j_{1}+k_{1}N_{\phi_{1}} & =\frac{\ell^{2}}{\ell_{1}^{2}}A_{k}-t_{1}t_{2}Q  Y_{s}\nonumber \\
j_{2}+k_{2}N_{\phi_{2}} & =\frac{\ell^{2}}{\ell_{2}^{2}}A_{k}+t_{1}t_{2}Q  Y_{s},
\end{align}
 where $A_{k}$ depends only on $k$ and $\beta$, but not on $s$,
while $Y_{s}$ depends only on $s$ and $\beta$, but not on $k$:
\begin{align}
A_{k} & =j_{1}+j_{2}+kN_{\phi}+\beta t_{1}Q\,.\nonumber \\
Y_{s} & =s-\frac{t_{2}j_{1}-t_{1}j_{2}+\beta t_{1}t_{2}Q}{t_{1}t_{2}N_{\phi}}\,.\label{eq:A and Y}
\end{align}
Using these definitions we see that the $z$-dependent factor on the
right-hand-side of Eq. \eqref{eq:app prod of phi's} does not depend
on the summation index $s$ and becomes rather simple: 
\begin{align}
e^{-2\pi i(j_{1}+k_{1}N_{\phi_{1}}+j_{2}+k_{2}N_{\phi_{2}})z} & =e^{-2\pi iA_{k}z}.\label{eq:phase factor}
\end{align}

Let us now consider the factor that is exponential in the winding
number. Using \eqref{eq:A and Y} it can be rewritten as 
\begin{multline}
\exp\left[i\pi\tau\left(\frac{(j_{1}+k_{1}N_{\phi_{1}})^{2}}{N_{\phi_{1}}}+\frac{(j_{2}+k_{2}N_{\phi_{2}})^{2}}{N_{\phi_{2}}}\right)\right]\\
=\exp\left[ i\pi\tau\left(\frac{A_{k}^{2}}{N_{\phi}}+t_{1}t_{2}N_{\phi}Y_{s}^{2}\right)\right]
\end{multline}
 i.e. it factorizes into two parts, each of which only depends on one
of the summation indices.

Most of the complication lies in the Hermite polynomials, at least
if $n_{1}\neq0,1$. The Hermite polynomial in Eq. \eqref{eq:app prod of phi's}
can be written as 
\begin{multline}
H_{n_{1}}\left(\frac{2\pi\ell_{1}}{L_{x}}(j_{1}+k_{1}N_{\phi_{1}})-\frac{y}{\ell_{1}}\right)  \\ =H_{n_{1}}\left(\frac{\ell}{\ell_{1}}\left[\frac{2\pi\ell}{L_{x}}  A_{k}-\frac{y}{\ell}\right]-\left(\frac{2\pi\ell_{1}}{L_{x}}t_{1}t_{2}Q\right)Y_{s}\right).
\end{multline}

 With the following identities
\begin{align}
H_{n}(x+y) & =\sum_{k=0}^{n}\binom{n}{k}H_{k}(x) (2y)^{n-k}\nonumber \\
H_{k}(\gamma x) & =\sum_{i=0}^{\lfloor k/2\rfloor}\gamma^{k-2i}(\gamma^{2}-1)^{i}\binom{k}{2i}\frac{(2i)!}{i!}  H_{k-2i}(x)\label{eq:hermite identity 1}
\end{align}
 we find that 
\begin{multline}
H_{n_{1}}\left(\frac{\ell}{\ell_{1}}\frac{2\pi\ell}{L_{1}}A_{k}-\alpha_{1}Y_{s}\right)\\
=\sum_{l=0}^{n_{1}}\binom{n_{1}}{l}\left(-2\alpha_{1}Y_{s}\right)^{n_{1}-l} H_{l}\left(\frac{\ell}{\ell_{1}}\frac{2\pi\ell}{L_{1}}A_{k}\right)\\
=\sum_{l=0}^{n_{1}}\sum_{i=0}^{\lfloor l/2\rfloor}\frac{(2i)!}{i!}\left(\frac{\ell}{\ell_{1}}\right)^{l-2i}\left(-\frac{\ell^{2}}{\ell_{2}^{2}}\right)^{i}\binom{l}{2i}\binom{n_{1}}{l}\\
\times \left(-2\alpha_{1}Y_{s}\right)^{n_{1}-l} H_{l-2i}\left(\frac{2\pi\ell}{L_{1}}A_{k}\right)
\end{multline}
 where $\alpha_{1}=2\pi\ell_{1}t_{1}t_{2}Q/L_{1}$ and $\lfloor k/2 \rfloor=k/2$ if $k$ is even, resp. $(k-1)/2$ if $k$ is odd. 
 The first line in Eq. \eqref{eq:hermite identity 1} can easily be derived using the generating function of the Hermite polynomials.  The second is more tedious to derive and follows from successive partial integration of the integral $\int dx \exp[-x^2] H_n(\gamma x) H_m(x)$. 
 
 We can now identify the coefficient $C_{j_{1},j_{2};j}^{n_{1}0;n}$ to be 
\begin{widetext}
\begin{multline}
C_{j_{1},j_{2};j}^{n_{1},0;n}  =\frac{\mathcal{N}_{n_{1}}^{\ell_{1}}\mathcal{N}_{0}^{\ell_{2}}}{n!\mathcal{N}_{n}^{\ell}}\sqrt{\frac{t_1}{t}}^n\sum_{i=0}^{\lfloor(n_{1}-n)/2\rfloor}\frac{n_1!}{(n_1-n-2i)!i!} 
  \left(-\frac{t_{2}}{t}\right)^{i}\left(-4\pi t_{1}t_{2}\sqrt{\frac{\Im(\tau)Q}{2\pi t}}\right)^{n_{1}-n-2i}\\ \times
  \sum_{s=-\infty}^{\infty}\left(s-\frac{t_{2}j_{1}-t_{1}j_{2}+\beta t_{1}t_{2}Q}{t_{1}t_{2}N_{\phi}}\right)^{n_{1}-n-2i} \exp\left[i\pi\tau t_{1}t_{2}N_{\phi}\left(s-\frac{t_{2}j_{1}-t_{1}j_{2}+\beta t_{1}t_{2}Q}{t_{1}t_{2}N_{\phi}}\right)^{2} \right]\:
\end{multline}
\end{widetext}
 where $j=(j_{1}+j_{2}+\beta t_{1}Q)\mbox{mod }N_{\phi}$. The coefficients vanish for other values of $j$. 
Equation \eqref{eq: Clebsch Gordan m00} is obtained by setting $n=0$. 
In order to find Eqs. \eqref{eq:Clebsch000} and \eqref{eq:Clebsch100} one must do a Poisson resummation on the sum over $s$.

%%%%%%%%%%%%%%%%%%%%%%%%%%%%%%%%%%%%%%%%%%%%%%%%%%%%%%%%%%%%%%%%%%%%%%%%%%%%%%%%%%%%%%%%%%%%%%%%%%%%%%%%%%%%%%%%%%%%%%%%%%%%%%%%%%%%%%%%%%%%%%%%%%%%%%%%%%%%%%%%%%%%%%%%%%%%%%%%%%%%%%%%%%%%%%%%%%%%%%%%%%%%%%%%%%%%%%%%%%%%%%%%%%%%%%%%%%%%%%%%%%%%%%%%%%%%%%%%%%%%%%%%%%%%%%%%%%%%%%%%

\end{document}